\documentclass[12pt]{article}
\usepackage{graphicx}
\usepackage{pict2e}
\usepackage{dcolumn}
\usepackage{bm}
\usepackage{comment}
\usepackage{rotating}
\usepackage{longtable}
\usepackage{float}
\usepackage{eucal}
\usepackage{csquotes}
\usepackage{xcolor} 
 
\usepackage{authblk}
\usepackage{longtable}

\makeatletter

\newcommand{\Rmnum}[1]{\expandafter\@slowromancap\romannumeral #1@}
\makeatother

\begin{document}
  

\title{Ultra-fast timing detectors to probe exotic properties of
nuclei using RIB facility}


\author[1]{Ushasi Datta \thanks{ushasi.dattapramanik@saha.ac.in}}
\author[1]{S. Chakraborty}
\author[1]{A. Rahaman}
 
\affil[1]{Saha Institute of Nuclear Physics, Kolkata 700064, India}

\date{\today}

\maketitle

\begin{abstract}
Recently,  the facilities of radioactive ion beam (RIB) combined with advanced detector systems provide us
 unique opportunity to probe the exotic
properties of the nuclei with unusual neutron-to-proton ratio.  In this article, a study of  
 characterization of  different types of ultra-fast timing detectors:  
a special type of gas detector (multi-strip multi-gap resistive plate chamber, MMRPC) ($\sigma$ $<$100 ps), 
 scintillators array ( viz., $LaBr_3:Ce$) (timing resolution ($\sigma<$250 ps)  are being presented. 
 A brief discussion on usage of these different types of ultra-fast 
timing detector systems  at radioactive ion beam facilities is also included.  
\end{abstract}


 \vspace*{-0.3cm}
\section{Introduction}
\vspace*{-0.2cm}
In recent decades, the-state-of-art  ultra-fast timing detectors  have evolved 
very rapidly toward ever-higher performance. These detectors  have few  key characteristics 
that enable their application to  expand our knowledge on basic and applied sciences with unique dimensions.
The  fast rise time ($<$10 ns) pulses  due to response of  the detector allow high-resolution measurements 
in the time-domain.  Moreover, due to the simplicity of detector construction, 
easy to handle and long time performance reliability, researchers are attracted for various  utilization of 
these types detectors. In this article, studies of the response of different types of ultra-fast timing
 detector systems  for detecting particles and $\gamma$-rays are presented. 
These  different types of ultra  fast timing detectors are special type of gas detector 
(multi-strip multi-gap resistive plate chamber, MMRPC) and scintillators ( viz., $LaBr_3:Ce$). 
Different types of  scintillator detectors  are commercially available and detector response mechanism is similar. 
Though depending upon light output and its decay constant, the utilities are different for optimum
uses. Ultra fast timing plastic scintillators array are used in many large-scale experiments for charge, time of flight
and position measurements of charged particles or nuclei. These detectors are also used for detecting neutrons. 
Due to fast decay time and high light output inorganic scintillators, like $LaBr_3$ etc. are very useful for detecting
$\gamma$-rays with good timing   and energy  resolution.  
 A brief discussion on usage of these detectors for studying the  properties of the exotic nuclei is also included.

\begin{figure*}[t]
\centering
\includegraphics[width=4.7cm,height=3.3cm]{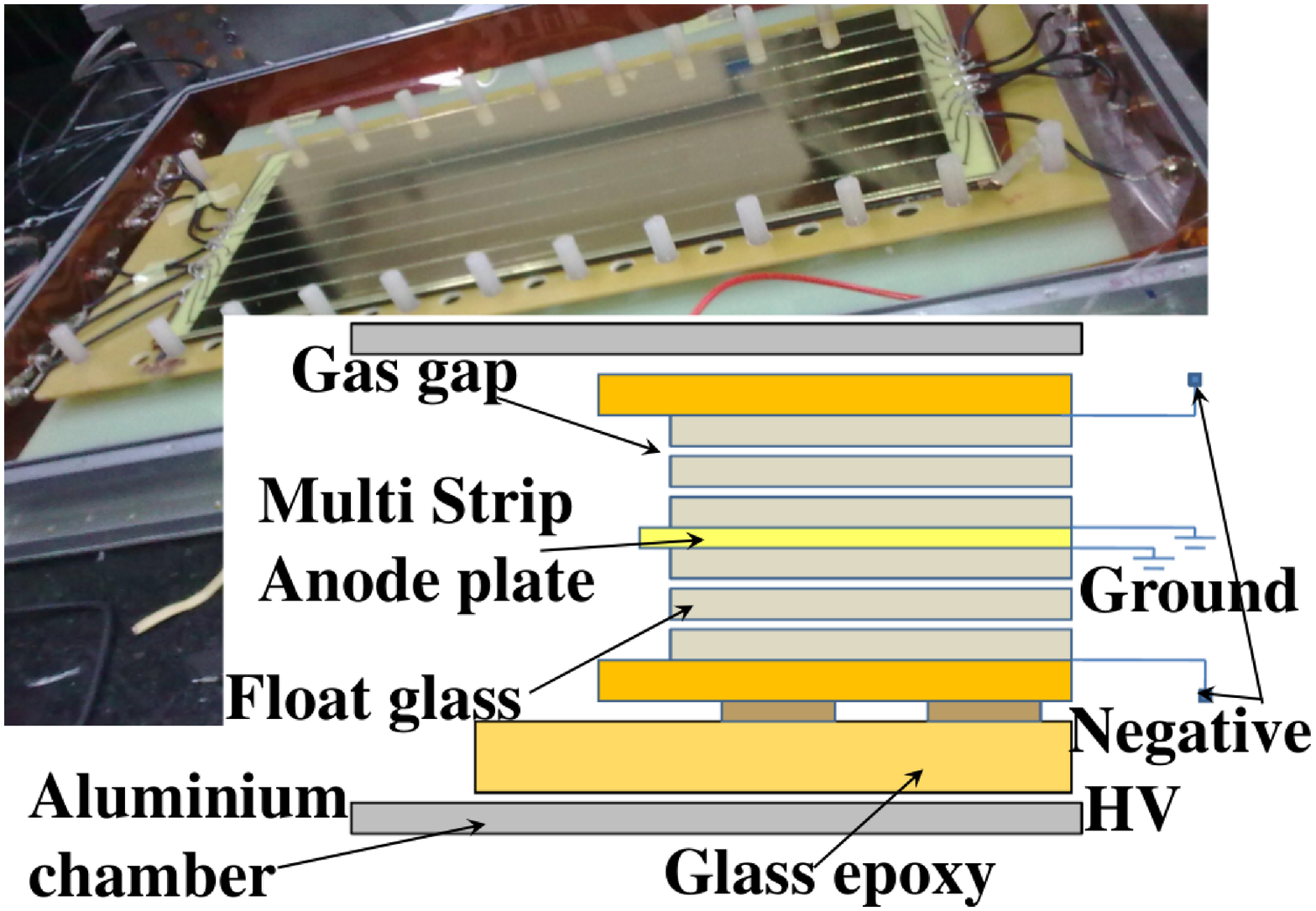}
\includegraphics[width=5.5cm,height=3.8cm]{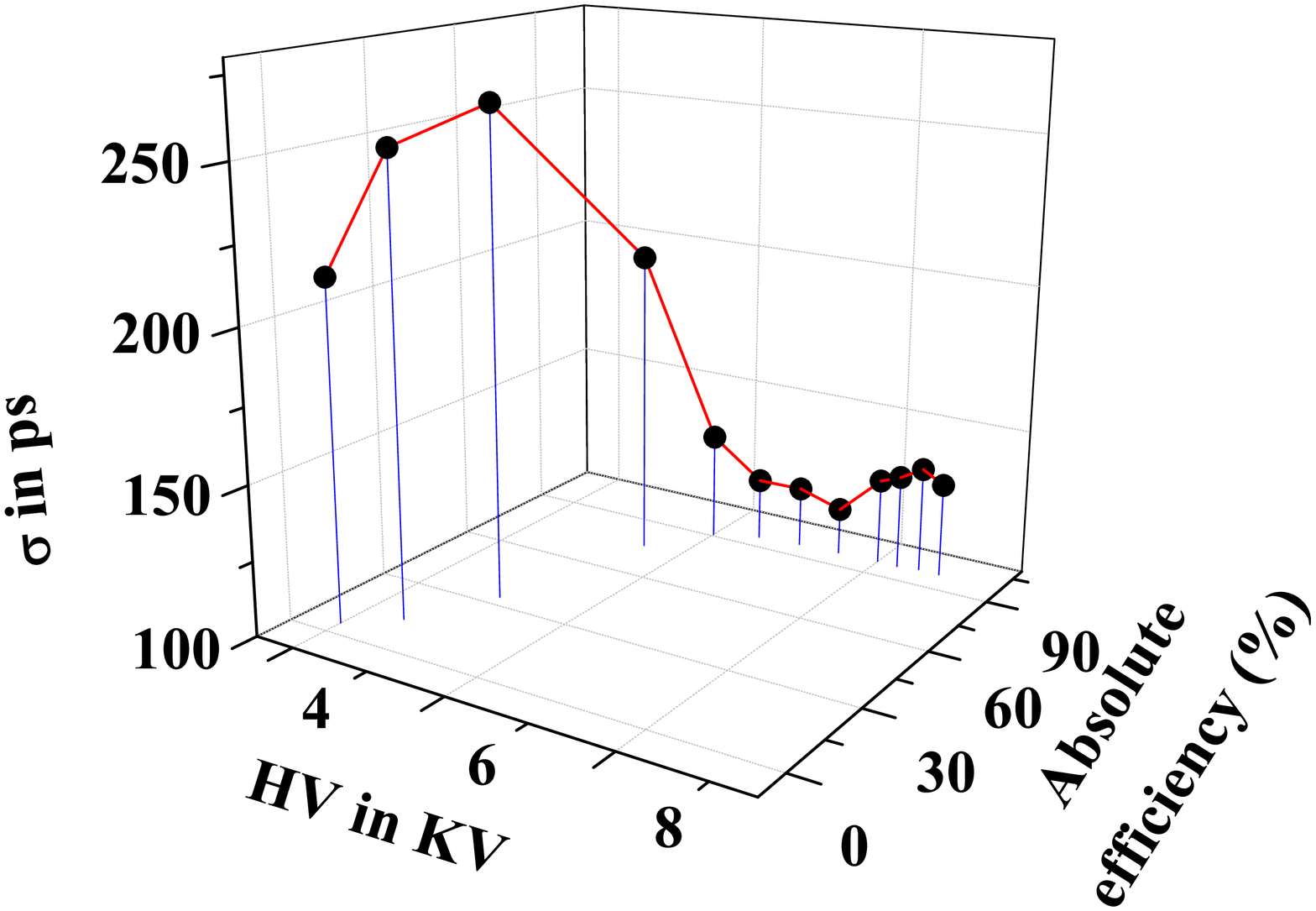}
\includegraphics[width=5.5cm,height=3.8cm]{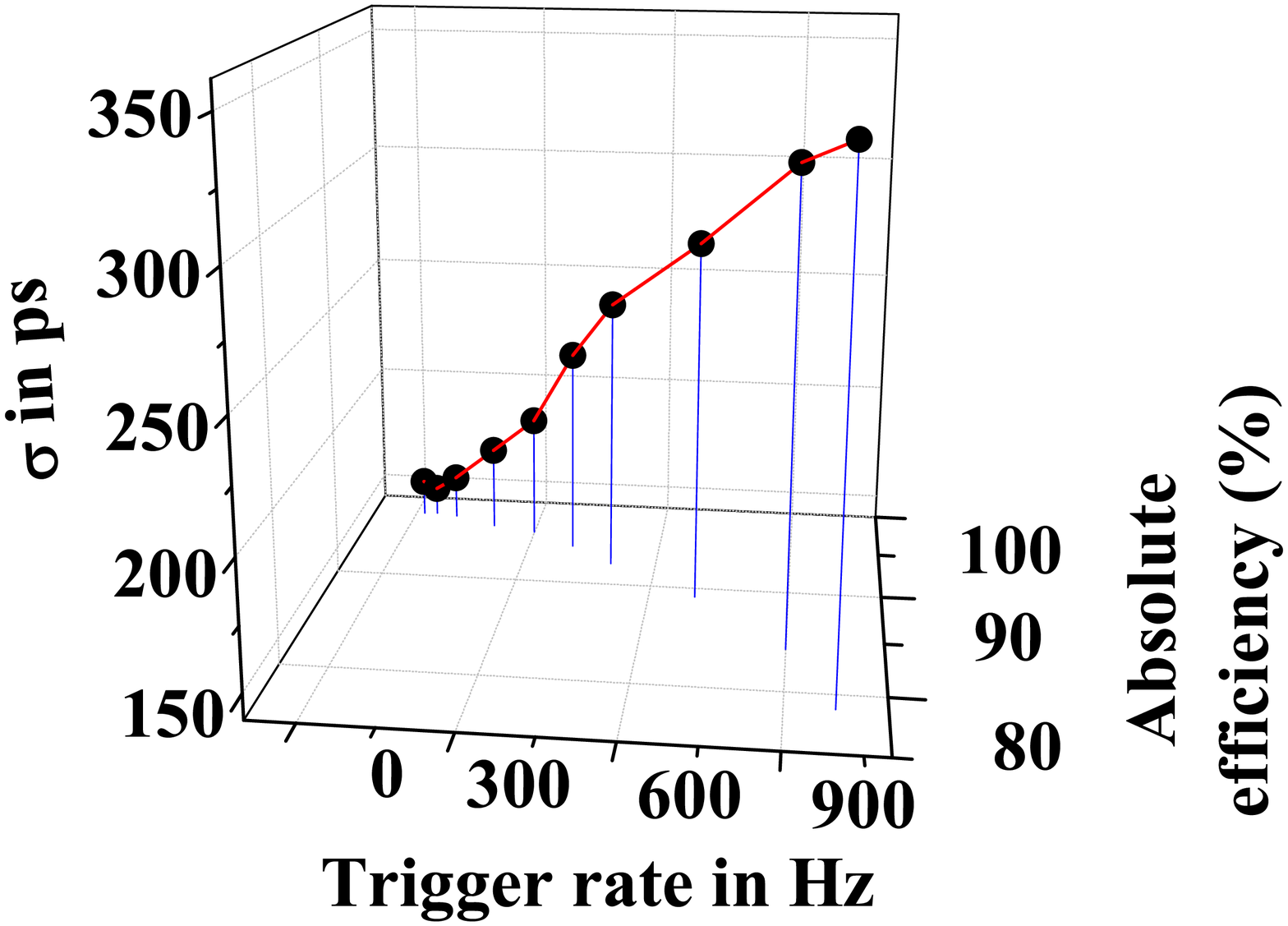}
\caption{ [Left] A top view photograph of a  Multi-strip Multi-gap Resistive Plate Chamber with the segmented anode strips. 
(inset) Sectional side view of the  developed Multi-strip Multi-gap Resistive Plate Chamber.  A three-dimensional 
plot of the time resolution of the MMRPC (after slew correction) against absolute efficiency,  applied bias 
voltage (middle) and  trigger rate (right)}
\label{rpc}      
\end{figure*}     	 
\vspace*{-0.3cm}
 
\section{Multi-strip multi-gap Resistive Plate Chamber (MMRPC) developed at SINP, Kolkata}
\vspace*{-0.2cm}

\begin{figure}[h]
\centering
\includegraphics[width=8cm,clip]{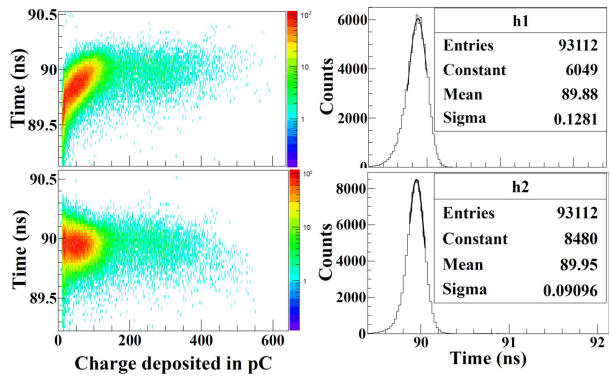}
\caption{Time distribution of MMRPC before (top-left) and after (bottom-left) slew correction,
 against the deposited charge on a strip for the events hitting only one strip at a time. 
(top-right)Time of MMRPC (before slew correction)  and (bottom-right) the same  after slew correction}
\label{slewd}       
\end{figure}

\begin{figure}[h]
\centering
\includegraphics[width=6.0cm,height=3.0cm]{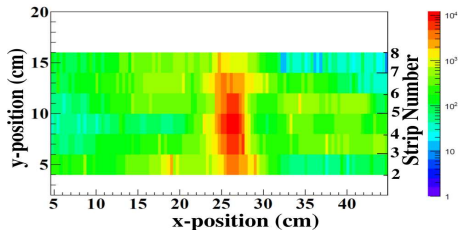}
\caption{Reconstructed image of electron beam spot at ELBE, obtained by present developed
detector via timing measurement.}
\label{position}       
\end{figure}

Three decades ago the Resistive Plate Chamber (RPC) \cite{santonico,cardarelli} was invented to overcome several
problems of parallel plate chambers. The working principle is similar to a gas detector in avalanche or streamer 
mode with modification due to  resistive plate cathode. The electrodes of RPCs are made
of resistive material like Bakelite or glass. This has the effect that only a limited part of the electrode
is discharged during the passage of an ionizing particle with subsequent avalanches or streamers,
while the rest of the electrode remains unchanged. To improve timing resolution, Multi-gap Resistive 
Plate Chamber \cite{zeballos,fonte} (MRPC) is an intelligent modification of a RPC by increasing the electric
field across the gas gap and the thickness of the gas gaps are reduced by inserting (electro-statically) floating
glasses between anode and cathode. Moreover, via segmentation of readout board,  high granularity can be achieved 
for a single MMRPC detector. Thus the MRPCs have  high granularity, high-resolution inexpensive TOF system 
(compared to standard scintillator with PMTs). This includes both
fundamental research in particle physics\cite{prpc}, astrophysics, 
cosmology, nuclear physics\cite{hades}, and applied research in medical imaging (cost effective Positron 
Emission Tomography)\cite{blanco,borozdin}, security purpose like  cosmic muon tomography, climate change etc. 
But, utilization of such detector in low-energy nuclear physics study
is scarce. To explore new applications, a prototype detector was developed at 
SINP, Kolkata \cite{udp12,udp15}. The prototype is a double stack glass MMRPC of size 40 cm $\times$ 20 cm
 with segmented anode strips (Figure \ref{rpc}-left). The cathode plates were made by introducing a thin 
layer of conducting  material on 1 mm thick float glass. 
Negative high voltage was supplied to the cathode while the anodes were kept at ground potential. 
A  non-flammable gas mixture of 89$\%$ R134a ($C_2H_2F_4$), 4.0$\%$ Sulfur Hexafluoride ($SF_6$) 
and 7.0$\%$ Isobutane ($C_2H_{10}$) was kept in flow mode at atmospheric pressure throughout the volume of the detector. 
The detailed response of the  developed detector  for  detecting cosmic muons and  $\gamma$-rays from 
radioactive source had been studied \cite{udp12}.   
For response of non-charged particle the charged clouds are produced in the counting gas via indirect processes 
and hence, the response time of the  MMRPC for the neutrons and $\gamma$-rays is worse in comparison 
to minimum ionizing particles (MIPs). 
The response of the  MMRPC  to cosmic muons and $\gamma$-rays using radioactive source ($^{60}Co$) have 
been studied in coincidence with fast scintillator, cerium doped Lanthanum bromide (LaBr$_3$:Ce) scintillators. 
The measured time resolution of the MMRPC for $\gamma$-rays is
 $\sim$270 ps which is much worse  in comparison to high energy cosmic muons (150 ps) as shown 
in Figure 4 (bottom-right) of Ref. \cite{udp12}. Subsequently, a more detailed  study of the MMRPC
detector was performed using a pulsed electron beam (pulse width $<$ 10 ps) of 29 MeV 
at  the ELBE facility \cite{udp15}. 
A sufficiently  long plateau has been observed ($\sim 95\%$) as 
shown in   Figure \ref{rpc}-middle. The time resolution of MMRPC improves with increasing negative
 bias voltage and reaches an optimum at 7.5 kV which is considered as the operating voltage. 
Figure \ref{slewd} (top-left) and (bottom-left) show the time against deposited charge in a 
single strip of the  MMRPC detector before and after slew correction, respectively. The measured 
time resolution ($\sigma$) of the  MMRPC for events hitting only a single strip after slew correction 
is 90$\pm$1 ps. The measured position  resolution of the MMRPC along and across the strip 
are 2.8$\pm$0.6 cm and 0.58 cm, respectively. Figure \ref{position} shows the reconstructed 
image of electron beam spot in the XY-plane of the MMRPC detector. 
The   advantage of the  detector is   fast timing response along with 
position resolution.  Other  practical advantages of MMRPCs are simple and easy of construction, operated  
in atmospheric gas pressure, working condition independent of temperature, negligible 
distortion in presence of magnetic field, cost-effective and possibility of utilization in large scale. 
The major limitation of this type of detector is lack of high event rate capability.
As shown in the Figure \ref{rpc}-right, the efficiency and timing resolution of 
the developed detector became worse with increasing  events rate. 
Considering above-mentioned advantages and limitations, these type of excellent detectors can be used
as a TOF for  relativistic energy neutrons\cite{alti,zolthan,jamil,wang}, protons\cite{sch09}, 
fragments/light charged particles\cite{diego}, 
and electrons\cite{udp15} as reaction products.  Detailed simulation work for response of
 MMRPC detector to detect low energy neutron has been reported by 
a number of research groups \cite{jamil}. We are planning to study the neutron response  
  using 14 MeV neutron generator\cite{bishnoi} with  a  modification of MMRPC. 
Now,  few examples for utility of MMRPC to study exotic nuclei  is being discussed.
The study of exotic heavy   nuclei (A $>$ 50) via invariant mass analysis is restricted  due to 
limitation of acceptance of the neutrons detection in the setup at various RIB facilities.  
Plastic scintillator based neutron detector of  larger dimension for better acceptance
 becomes commercially inviable. The low cost MMRPC based large area neutron detector can be the  
alternative to provide neutron detection with good time and position resolution. 
Study of the exotic nuclei beyond drip  line by proton knockout of neutron-rich nuclei is an excellent tool. 
 But the measurements may be  affected due to reaction product of other  channels.
 The tagging of knockout proton using this detector may solve the problem 
and  provide valuable information about the  unbound nuclei.

\vspace*{-0.5cm}

\vspace*{-0.5cm}

\begin{figure}[h]
\centering
\includegraphics[width=3.5cm,height=2.0cm]{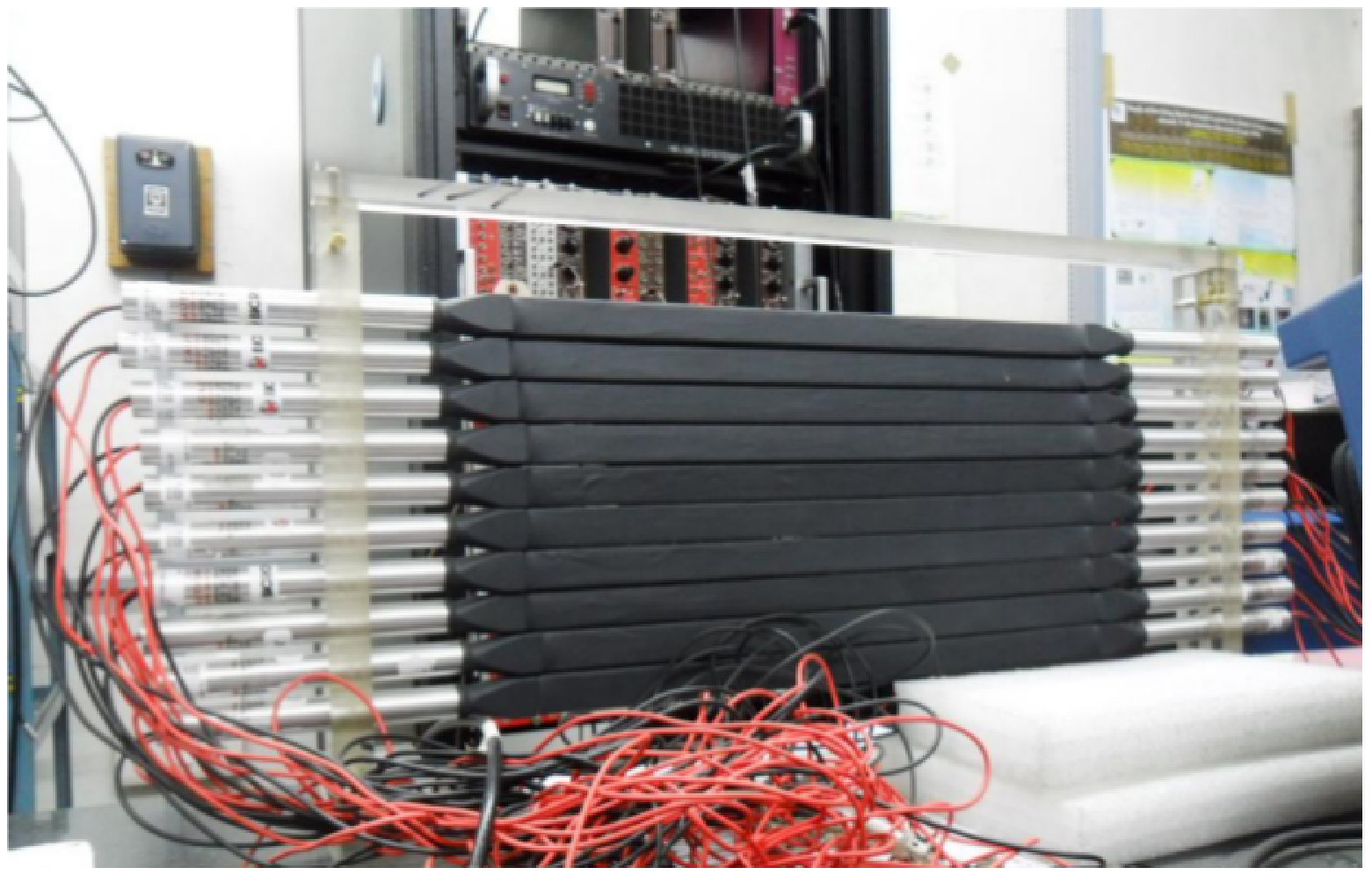}
\includegraphics[width=3.5cm,height=2.0cm]{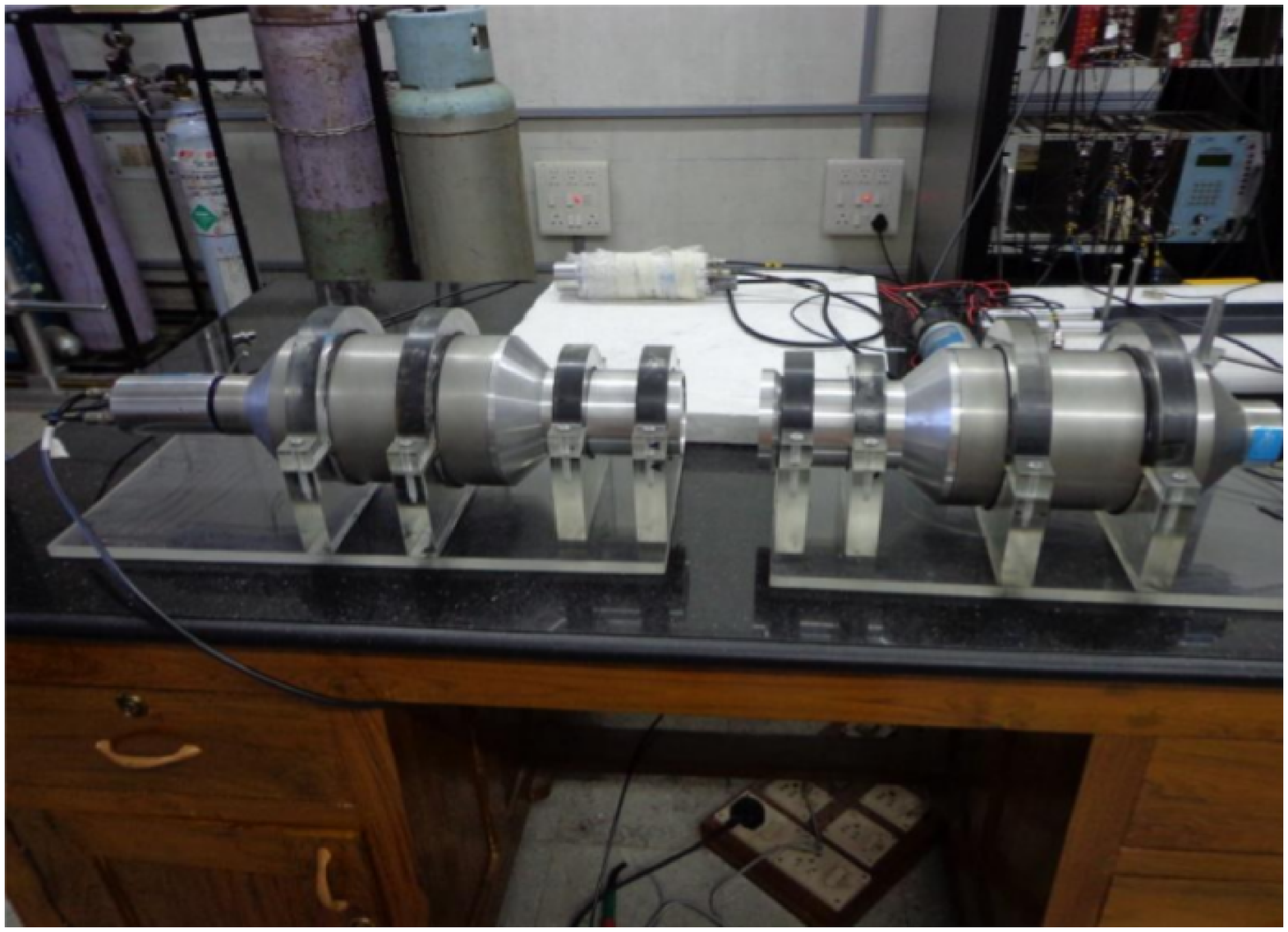}
\caption{ Photograph of the plastic scintillator array(left) and LaBr$_3$:Ce detectors(right).}
\label{plastic}       
\end{figure}
\vspace{-0.4cm}
\section{Scintillators}
\vspace*{-0.2cm}
\subsection{$LaBr_3(Ce)$}
Lanthanum Bromide, cerium doped, $LaBr_3(Ce)$ is one of the new generation inorganic scintillator for detecting $\gamma$-rays 
which  exhibit  improved energy resolution, faster decay time, and temperature stability compared to 
available other inorganic scientillators; NaI, CsI etc. Figure \ref{plastic}-right shows the photograph of LaBr3 detector
of six inches long crystal with fast PMTs. A detailed response of this types of detector with various
 dimensions and readout systems has been studied using using standard $\gamma$-ray sources 
($^{60}$Co, $^{152}$Eu, $^{137}$Ba) and cosmic background \cite{udpdae,udp12} at SINP laboratory.
It has been observed that the energy response of   $\gamma$-rays using this crystal is highly nonlinear (see figure 5)
and  the energy  resolution is inversely proportional to the square root of the energy ( $R(E)\propto 1/E^{1/2}$).
Figure 5 [right] shows timing spectra (t$_1$-t$_2$)  and inset shows the energy spectra of $^{60}$Co source.
The dashes and solid  line  represents overall TAC (time-to-amplitude converter) and photo-peak projected TAC spectra. 
Measured time resolution is better for the smaller crystal (2.5 cm diameter and length) ($\sigma \sim$ 120 ps)\cite{udp12} compared to the 
larger crystal ($7.6$ cm dia and $15.6$ cm length) ($\sigma \sim$ 225  ps). Due to fast timing response compared to HpGe and NaI(Tl) 
detectors, the LaBr3 detectors are useful in the experiment that require good timing resolutions, such as   
 lifetime measurements of isomeric states in the nuclei and one may pin down various structural properties of exotic nuclei \cite{reg}. 
This detector will be particularly useful for the  measurement of life time  which are not 
accessible by Doppler Shift Attenuation  (DSA) and plunger method.
Due to availability of larger dimensions of this type crystal, high energy $\gamma$-rays  
can be detected with better photo peak efficiency which may be helpful for measurements of the low cross section specific events
 of the nuclei. Study of the ground state configuration of the exotic nuclei by Coulomb breakup,  knockout reaction at intermediate energy  (E $>$ 100 AMeV)  
are affected by identification of core excited states. This type of detectors with better energy resolution along with efficiency,
may provide excellent solution.  However, the plan of the experimental setup  should be such that the Doppler broadening effect should 
not suppress the intrinsic detector resolution.
Using this advanced detector,  measurements of p-$\gamma$ branching ratio, relevant to nuclear astro-physics is another
important utility.

\begin{figure*}[t]
 \includegraphics[width=5.0cm,height=4.0cm]{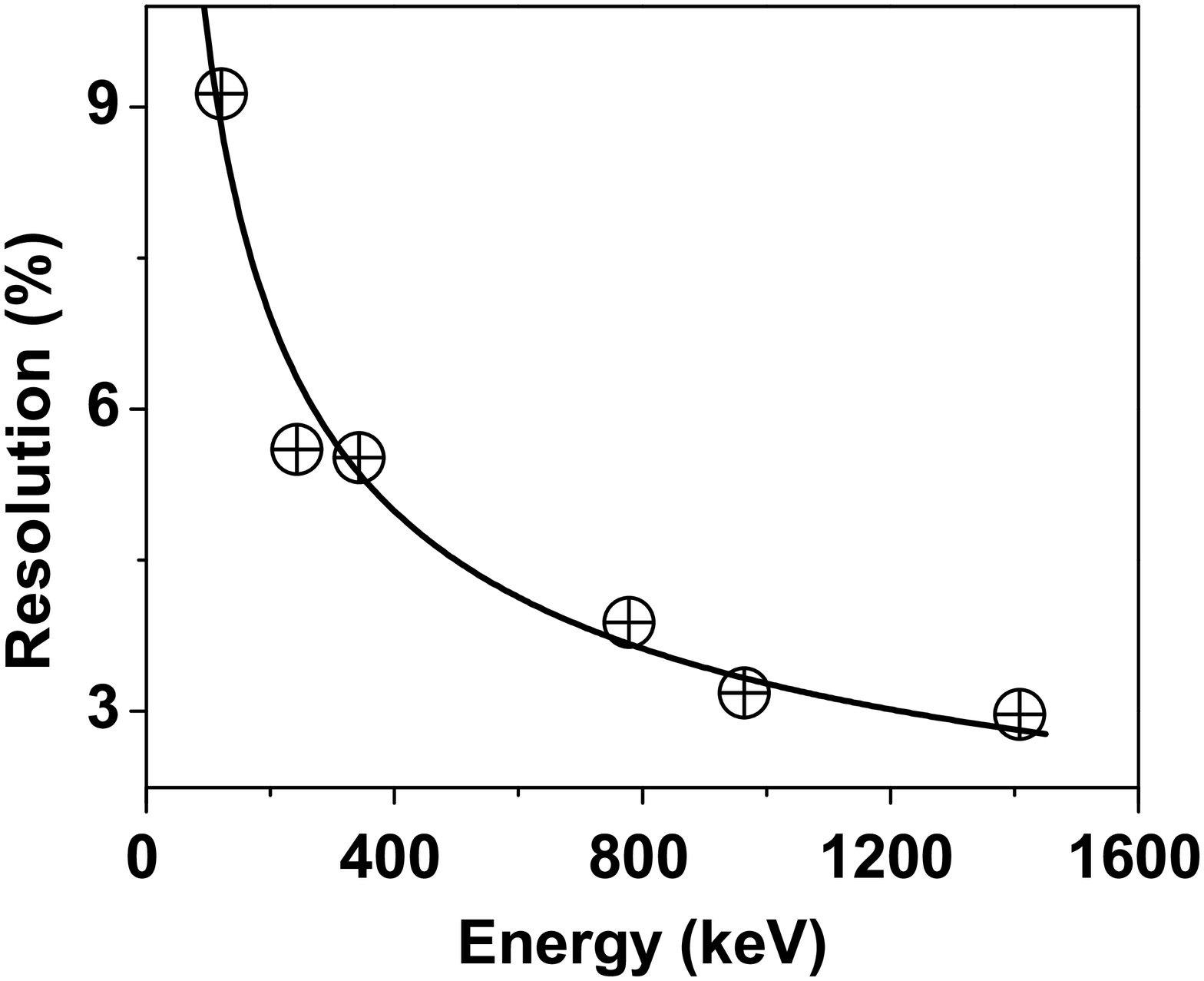}
 \includegraphics[width=5.0cm,height=4.0cm]{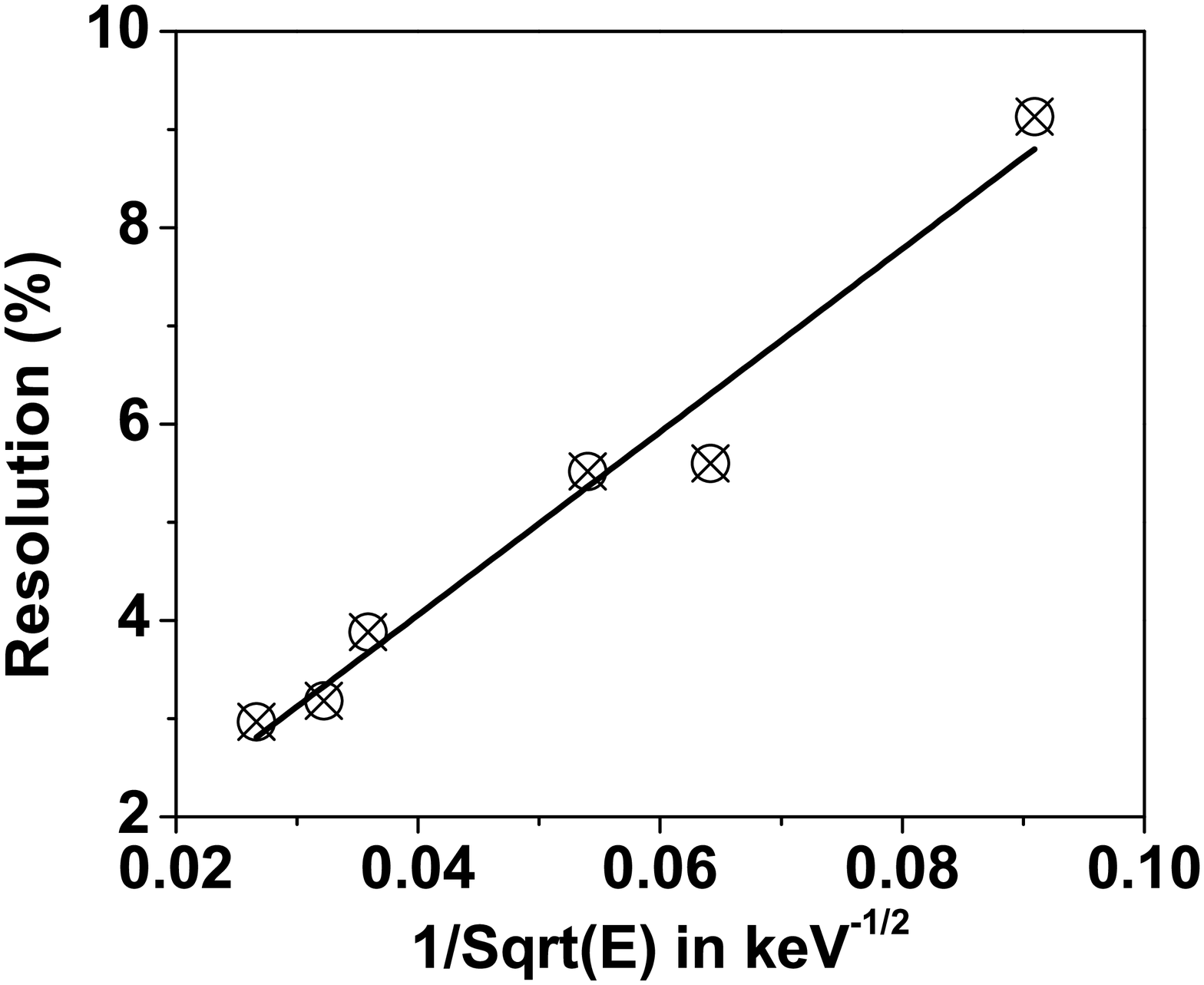}
 \includegraphics[width=4.8cm,height=3.9cm]{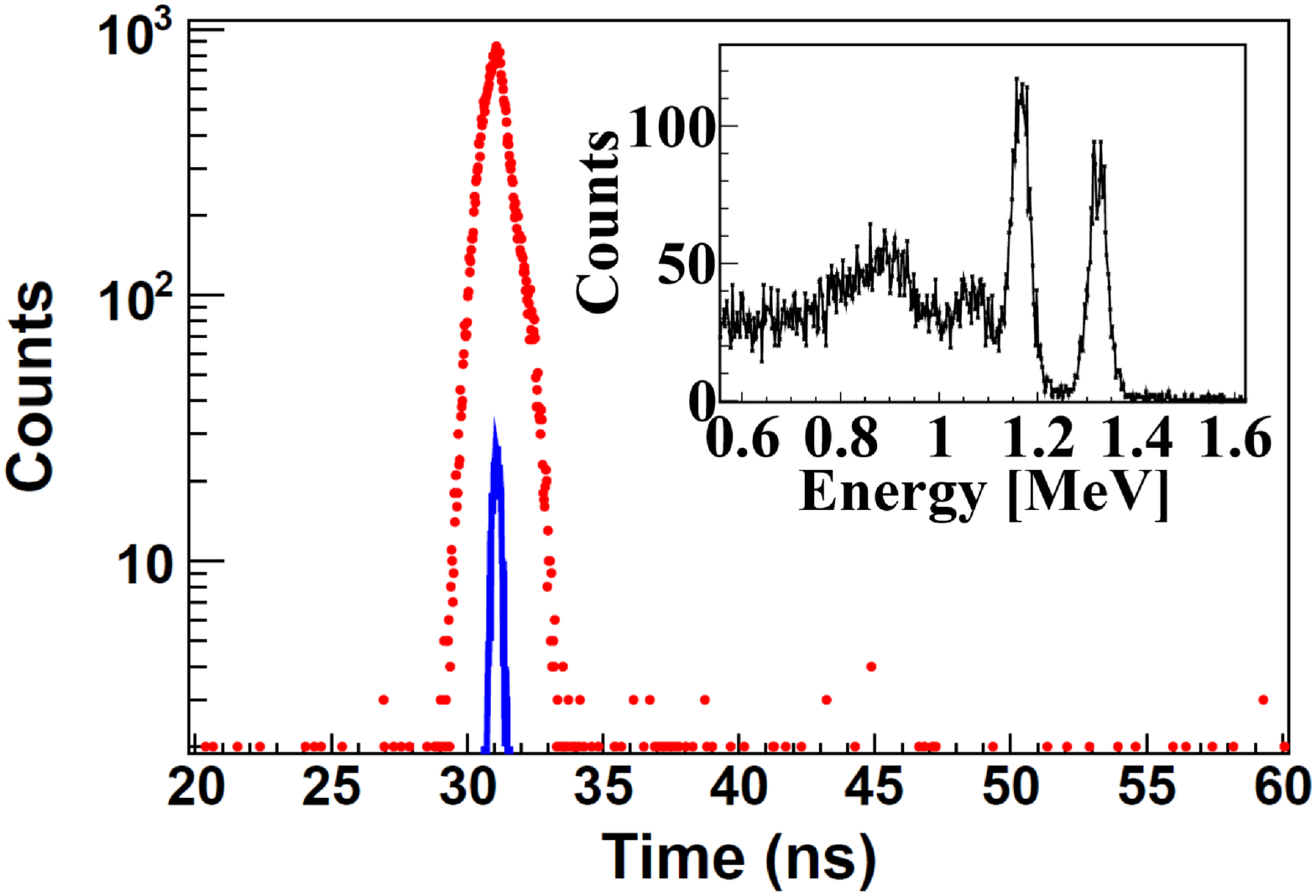}
 \caption{[left] Energy resolution for detecting $\gamma$-rays 
 against  energy  of the $\gamma$-rays ($^{152}$Eu source)
for lanthanum bromide detector. [middle] The functional dependency 
of energy resolution of  detected the $\gamma$-rays. 
[right] TAC timing spectra obtained for  $\gamma$-rays 
($^{60}$Co) in coincidence. The energy spectra of $\gamma$-rays are shown in the inset. } 
\label{gamma}
\end{figure*}
\vspace*{-0.0cm}

\subsection{Plastic scintillator}
\vspace*{-0.1cm}
Plastic scintillator detectors are widely used for  measurement of the time of flight and charges  
 of high velocity  particles  in many experiments of nuclear \cite{rah13}
and high energy physics. In house facility at SINP offers us to measure various intrinsic
 properties of those detector like measurement of rise time, the optical attenuation length of the
 material \cite{rahdae} etc.. Figure \ref{plastic}-left shows the photographs of plastic scintillator 
 array with equivalent size and granularity of the detection channels of our developed MMRPC.
 Though, this detector as TOF  is expensive compared to MMRPC, but
 considering the  capability of high events rates, these types of detectors are being used in 
wide range of experiment using exotic nuclei with high intensity. It would be interesting to
 explore composite detector TOF using MMRPC and plastic scintillator.
\vspace*{-0.4cm}      
\section{Applications of the ultra-fast timing detectors to the society}
\vspace*{-0.3cm}
The applications of ultrafast technology, are wide in horizon. Here we are restricting 
our views with above-mentioned detectors. Use of cost-effective MRPCs in PET technology\cite{blanco,david,amaldi} 
for  medical imaging will be a great use of society. Positron Emission Tomography (PET)
imaging technique is  based on positron emitting radio-tracer and detecting the two back-to-back
 $\gamma$-rays emitted when the positron annihilates in the patient’s body.  A
line is drawn connecting the detected positions of the two $\gamma$-rays; thus an
image can be constructed when many such lines have been recorded.
If the time of arrival of the $\gamma$-rays in the detector can be measured (with 
precision), then the position of the positron annihilation can be localized along
this line. This technique is known as TOF-PET and studies have shown that the resultant
image becomes much clearer due to the reduction of the background. 
 Muon tomography\cite{borozdin} is a technique  based on the measurement of multiple scattering of 
atmospheric cosmic muons. This is a promising technique for detecting 
 and imaging  heavily shielded high-Z nuclear materials{\cite{wang15}. This technology will 
enables large-scale imaging of cargos, trucks etc.  at border crossing  or  airport  
penetrating thick layers of steel. 
 \vspace*{-0.4cm}
\section{Summary}
\vspace*{-0.3cm}
Detailed response of  different types of ultra-fast timing detectors; a special type of gas detector,  MMRPC
($\sigma$ $<$100 ps),  scintillators array  ( viz., $LaBr_3:Ce$) ($\sigma$ $<$250 ps) 
have been discussed. A wonderful, reconstructed 
image of the electron beam spot provide unique  capability  of imaging  of the MMRPC  which are  almost two order of 
magnitude cheaper than scintillators.  
But  due to limitation in rate handling capability, this detector can be used for detecting low rate reaction products. For larger 
rate handing capability, plastic scintillator are the uniqe  for detecting purpose.
Due to fast decay time and high light output inorganic scintillators, like $LaBr_3$ etc. are very useful for detecting
$\gamma$-rays with good timing ($\sigma \sim$ 150-250 ps) and energy 
resolution ($\Delta E/E$ $\sim3\%$)\cite{udpdae,udp12,reg}. We have studied the response of the charge particle of this type detector\cite{udp12}.
\vspace*{-0.4cm}
\section{Acknowledgement} 
\vspace*{-0.3cm}
 Authors are deeply thankful to DAE, India; workshop of SINP, Kolkata and  ELBE facility, Dresden  for their support.     

\end{document}